\documentclass[useAMS,usenatbib]{mn2e}

\usepackage{graphicx}
\usepackage{dcolumn}
\usepackage{bm}
\usepackage{amsmath}

\title[High-energy emission as a test of the prior emission model for GRB afterglows]
{High-energy emission as a test of the prior emission model for gamma-ray burst afterglows
}
\author[Murase et al.]
{Kohta Murase$^{1}$\thanks{E-mail: kmurase@yukawa.kyoto-u.ac.jp},
Kenji Toma$^{2}$, 
Ryo Yamazaki$^{3}$, 
Shigehiro Nagataki$^{1}$,
and Kunihito Ioka$^{4}$
\\
$^{1}$Yukawa Institute for Theoretical Physics, Kyoto University, Sakyo-ku, Kyoto 606-8502, Japan\\
$^{2}$Department of Astronomy \& Astrophysics; Center for Particle Astrophysics, Pennsylvania State University, University Park, PA 16802, USA
\\
$^{3}$Department of Physical Science, Hiroshima University, Higashi-Hiroshima 739-8526, Japan
\\
$^{4}$Theory Division, KEK (High Energy Accelerator Research Organization), Tsukuba 305-0801, Japan
}
\begin{document}

\date{}
\pagerange{\pageref{firstpage}--\pageref{lastpage}} \pubyear{2009}
\maketitle
\label{firstpage}

\begin{abstract}
We study high-energy gamma-ray afterglow emission from gamma-ray bursts (GRBs) in the prior emission model, which is proposed to explain the plateau phase of the X-ray afterglow. This model predicts the high-energy gamma-ray emission when the prompt GRB photons from the main flow are up-scattered by relativistic electrons accelerated at the external shock due to the prior flow. 
The expected spectrum has the peak of $\sim 10$--100~GeV at around the end time of the plateau phase for typical GRBs, and high-energy gamma rays from nearby and/or energetic GRBs can be detected by current and future Cherenkov telescopes such as MAGIC, VERITAS, CTA, and possibly \textit{Fermi}.
%
%
Multi-wavelength observations by ground-based optical telescopes as well as \textit{Fermi} and/or \textit{Swift} sattelites are important to constrain the model.
Such external inverse-Compton emission may even lead to GeV-TeV gamma-ray signals with the delay time of $\sim 10$--100~s, only if the plateau phase is short-lived.     
\end{abstract}


\section{Introduction}
High-energy gamma-ray emission from gamma-ray bursts (GRBs) has been expected 
for long years since EGRET detected several GRBs with GeV emission \citep[e.g.,][]{Hur+94}. 
Recently, the \textit{Fermi} 
Large Area Telescope (LAT) has detected high-energy ($>$ GeV) gamma rays from a fraction of GRBs \cite{Abd+09a,Abd+09b,Abd+09c}. 
Those detections have given us not only clues to the prompt emission mechanism but also chances to test new physics such as the Lorentz-invariance violation predicted in quantum gravity theories \cite{Abd+09b}. High-energy gamma rays can also be expected for afterglows. High-energy gamma rays from GRB 080916C, 090510 and 090902B could be attributed afterglow emission rather than the prompt emission \cite{KD09,Ghi+09,Gao+09}. 

Since the launch of the \textit{Swift} satellite, early X-ray afterglow emission has been one of the mysteries. They are classified into three phases: the steep decay phase, the plateau phase and the normal decay phase \citep[see, e.g.,][and references there in]{Nou+06,OBr+06,Zha+06}. 
In particular, the plateau phase is difficult to be explained by the classical external shock model \citep[e.g.,][]{Mes06,Zha07}, and numerous models have been proposed so far \citep[e.g.,][]{EG06,Nou+06,ITYN06,TIYN06,Zha+06,Der07,GDM07,Ghi+07,Pan07,UB07}. 
Various theoretical possibilities of high-energy emission have also been discussed by numerous authors, and modified forward shock models, where electrons accelerated at the shock lead to the synchrotron self-Compton (SSC) and/or external inverse-Compton (EIC) emission, 
have often been considered \citep[see, e.g.,][and references there in]{FP08}. 

Recently, Yamazaki (2009) proposed an alternative model, 
in which the X-ray light curves can be explained by the difference between the X-ray onset time and the burst trigger time. This model is somewhat different from the prior activity model, where the interaction with the prior flow is important and the EIC emission can also be predicted \cite{ITYN06}. In this work, we consider consequences of the prior emission model on high-energy emission. We show that multi-wavelength observations provide a useful test. 

\section[]{The Prior Emission Model}
Here, we briefly describe the prior emission model \cite{Yam09}. This model is a kind of two-component emission model, where prompt and afterglow emission are attributed to radiation from the main and the prior outflows, respectively. 
The behavior of light curves during the plateau phase is explained by the difference between the afterglow onset time and the trigger time. The onset time is usually set to the trigger time $t_0$ when the prompt emission starts, that is, $t_0=0$ is assumed. However, if $t_0 \neq 0$, the afterglow flux is written as $F_{\rm AG} \propto t^{-\alpha_0}= {(T+t_0)}^{-\alpha_0}$.  
Since $F_{\rm AG}$ is constant for $T<t_0$ and $F_{\rm AG} \propto T ^{-\alpha_0}$ for $T>t_0$, 
the light curve well matches the observed plateau phase as well as the subsequent normal decay phase.
%
Here we assume that the prior outflow produces X-rays via external forward shock emission.
Such a forward shock model is supported by the lack of spectral evolution across the transition from the plateau to the normal decay phase and the compliance of the ``closure relations'' in the normal decay phase after the transition \cite{LZZ07}. However, it still has several problems in e.g., explanation of chromatic breaks, 
and avoiding too bright precursor emission \cite{Yam09}. 

In this model, the prior outflow is typically assumed to precede the main outflow producing the prompt emission, and then the afterglow is occurring during the prompt emission so that prompt photons are naturally up-scattered by electrons accelerated at the external shock as long as both jets are in the line of sight. The inverse-Compton emission is characterized by the EIC Compton Y parameter $Y_{\rm EIC}$. 
Typical long GRBs have the luminosity of $L_{\rm GRB}^b \sim {10}^{51.5}~{\rm erg}~{\rm s}^{-1}$ at the break energy of $E^b \sim 0.1$--1~MeV and the duration of $\Delta T \sim {10}^{1.5}$~s, and then the expected EIC luminosity is very roughly written as $L_{\rm EIC} \sim {\rm min} (\frac{Y_{\rm EIC} L_{\rm GRB}^b \Delta T}{{\Delta t}_{\rm EIC}}, L_e)$, where ${\Delta t}_{\rm EIC} \approx \frac{r}{2 \Gamma^2 c}|_{t=t_0} \sim t_0$ is the duration of the EIC emission \citep[e.g.,][]{Fan+08}. At $t_0$, by using the standard external forward shock model \citep[e.g.,][]{SPN98}, we also obtain the bulk Lorentz factor of the prior outflow as $\Gamma \simeq 44~\mathcal{E}_{k,53}^{1/8} n_0^{-1/8} t_{0,3}^{-3/8}$ and the external shock radius as $r \simeq 2.3 \times {10}^{17}~{\rm cm}~ \mathcal{E}_{k,53}^{1/4} n_0^{-1/4} t_{0,3}^{3/4}$, where $\mathcal{E}_{k}$ is the isotropic kinetic energy of the prior outflow and $n$ is the circumburst medium density. 
In the slow cooling case, $\gamma_{e,m} < \gamma_{e,c}$, $Y_{\rm EIC}$ can be approximated in the Thomson limit for $E< E_{\rm KN,1} \equiv \Gamma \gamma_{e,m} m_e c^2$ as $Y_{\rm EIC} (E) \equiv \frac{E F_{\rm EIC}(E)}{\bar{E} F_{\rm GRB} (\bar{E})} \sim \gamma_{e,m}^2 \tau_T$, where $\bar{E} \sim E/\gamma_{e,m}^2$, $\tau_T \sim (\sigma_T \mathcal{N}_e/4 \pi r^2)$ is the Thomson optical depth, and $\mathcal{N}_e$ is the number of electrons. 
However, the Klein-Nishina (KN) effect becomes important for $E > E_{\rm KN,1}$ and, in fact, we can see $\gamma_{e,m}^2 E^b \gg E_{\rm KN,1}$.  
Then, in our case, the EIC emission at $E > E_{\rm KN,1}$ is dominated by radiation from electrons with $\gamma_e \approx E/\Gamma m_e c^2$ 
interacting with prompt photons with the energy of $\bar{E} \approx \Gamma^2 m_e^2 c^4/E$ via the Thomson scattering. As a result, we have $Y_{\rm EIC} (E)\equiv \frac{E F_{\rm EIC} (E)}{\bar{E} F_{\rm GRB} (\bar{E})} \sim \frac{E^2}{\Gamma^2 m_e^2 c^4} \tau_{e}$, where $\tau_e  \sim \tau_T {(\gamma_e/\gamma_{e,m})}^{-p+1}$ for $\gamma_{e,m} \leq \gamma_e<\gamma_{e,c}$ and $\tau_e \sim \tau_T  {(\gamma_{e,c}/\gamma_{e,m})}^{-p+1} {(\gamma_e/\gamma_{e,c})}^{-p}$ for $\gamma_e \geq \gamma_{e,c}$, where $p$ is the spectral index of forward shock electrons. 
The resulting EIC spectrum is roughly expressed as \citep[see, e.g.,][for general discussions]{GG03,NAS09}
\begin{equation}
E_{\rm EIC} F_{\rm EIC} (E) 
\propto
\left\{ \begin{array}{ll} 
E^{2-\beta_l}
& \mbox{($E < E_{\rm KN,1}$)}\\
E^{\beta_l-p+1}
& \mbox{($E_{\rm KN,1} \leq E < E_{\rm KN,2}$)}\\
E^{\beta_l-p}
& \mbox{($E_{\rm KN,2} \leq E$)}
\end{array} \right.
\end{equation}
where $E_{\rm KN,2} \equiv \Gamma \gamma_{e,c} m_e c^2$ and 
$\beta_l \sim 1$ is the low-energy photon index of the prompt emission, 
and we have assumed the EIC cooling time by prompt photons, 
$t_{\rm EIC}$, is long enough (see below). 
The electron minimum Lorentz factor $\gamma_m$ is estimated as $\gamma_{e,m} \sim {10}^{3-4}$ at $t \sim {10}^{3}$~s for typical GRB afterglows, where we can expect $E_{\rm KN,1} \sim 10-100$~GeV and $L_{\rm EIC} \sim {10}^{49}~{\rm erg} ~{\rm s}^{-1}$. 
The EIC emission discussed here is similar to that originally proposed for flares \citep[e.g.,][]{Wan+06,Wei+06,YD09}. 
However, the KN effect is much more significant in our case, 
and the spectrum with the peak of $\sim 10$--100~GeV is different from 
that in flares. Note that the total radiation energy of the EIC 
emission cannot exceed the total electron energy input at the external shock
of the prior outflow.

We are interested in the case where the EIC flux exceeds the SSC flux of the afterglow. Therefore, we need to estimate the SSC flux to see whether the EIC emission is observable or not.    
The characteristic energies in the SSC emission are obtained as \citep[e.g.,][]{SE01,ZM01}
\begin{eqnarray}
E_{\rm SSC}^m &\simeq& 2.3 \times {10}^{3}~{\rm eV}~g_{-1}^4 
~{\epsilon}_{e,-1}^4~{\epsilon}_{B,-2}^{\frac{1}{2}}
~\mathcal{E}_{k,53}^{\frac{3}{4}}~n_0^{-\frac{1}{4}}
~t_4^{-\frac{9}{4}} \\
E_{\rm SSC}^c &\simeq& 2.2 \times {10}^{10}~{\rm eV}
~{\epsilon}_{B,-2}^{-\frac{7}{2}}~{\mathcal{E}}_{k,53}^{-\frac{5}{4}}
~n_0^{-\frac{9}{4}}~{(1+Y)}^{-4}~{t}_{4}^{-\frac{1}{4}},
\end{eqnarray}
where $\epsilon_B$ and $\epsilon_e$ are the fractions of the shock energy transferred to the downstream magnetic fields and non-thermal electrons, respectively, and 
$g(p)=\frac{p-2}{p-1}$ for $p>2$. 
For $\epsilon_e > \epsilon_B$, we have $Y \sim Y_{\rm SSC} \sim \sqrt{(\gamma_{e,m}/\gamma_{e,c})^{p-2} (\epsilon_e/\epsilon_B)}$ \citep[e.g.,][]{SE01,ZM01} unless the EIC cooling is more important than the synchrotron and SSC cooling. In the slow cooling case, the SSC spectrum is expressed as
$F_{\rm SSC} \propto {E_{\rm SSC}}^{1/3}$ for 
$E_{\rm SSC} < E_{\rm SSC}^{m}$, $F_{\rm SSC} \propto
{E_{\rm SSC}}^{-(p-1)/2}$ for $E_{\rm SSC}^{m} < E_{\rm SSC} < 
E_{\rm SSC}^c$ and $F_{\rm SSC} \propto
{E_{\rm SSC}}^{-p/2}$ for $E_{\rm SSC}^{c} < E_{\rm SSC} < 
E_{\rm SSC}^{\rm cut}$. 
Here $E_{\rm SSC}^{\rm cut}$ is the cutoff energy determined either by the pair-creation opacity or the KN limit \citep[e.g.,][]{ZM01}. Note that only the first SSC component is important, since the second SSC component is typically negligible due to the KN suppression in the optically thin synchrotron scenario. The energy flux at the SSC peak is evaluated as
\begin{eqnarray}
E_{\rm SSC}^c~F_{\rm SSC}^{c}~&\simeq&~9.7~\times~{10}^{-8}
~{\rm GeV}~{\rm cm}^{-2}~{\rm s}^{-1}~
\frac{Y_{\rm SSC}}{ {(1+Y_{\rm SSC})}^{p-1}} \nonumber \\
&\times&~ g_{-1}^{3-p} \epsilon_{e,-1}^{3-p} 
\epsilon_{B,-2}^{2-p} n_0^{\frac{2-p}{2}}
\mathcal{E}_{k,53}^{\frac{4-p}{2}} t_{4}^{-\frac{4-p}{2}} D_{28}^{-2},
\end{eqnarray}
by which we normalize the SSC spectrum. The temporal behavior would change after the jet break time of $t_{j} \sim {10}^{5}$~s \cite{Rhoads99}, but we focus on the behavior 
before $t_j$.  

\section[]{The Resulting Gamma-Ray Flux}
In this work, we shall use the following approximate formula in order to calculate the EIC emission. 
The EIC flux observed at $t$ is written as \citep[e.g.,][]{AA81,Fan+08,TWM09}
\begin{eqnarray}
F_{\rm EIC} (E) &\simeq& \frac{3}{2} \sigma_T (1-\cos \theta^{\prime}) \int 
d \gamma_e \frac{f_{\rm EIC}}{4 \pi r^2} \frac{d \mathcal{N}_e}{d \gamma_e} 
\int d y \, 
\nonumber \\
&\times& \frac{\bar{F}_{\rm GRB}^b |_{t_0} G (\varepsilon)}{{(1+\Gamma^2 \theta^2)}^2}
\left[ 1- 2 y +2 y^2 + \frac{\xi^2}{2(1-\xi)} \right]
\end{eqnarray}
where 
$y \equiv \frac{\xi m_e c^2}{2(1-\cos \theta^{\prime}) \gamma_e \varepsilon (1-\xi)}$ 
and 
$\xi \equiv \frac{(1+z) (1+ \Gamma^2 \theta^2) E}{2 \Gamma \gamma_e m_e c^2}$.
Scattering angles $\theta$ and $\theta^{\prime}$  of EIC photons 
are measured in the central engine frame and the comoving frame, respectively, and 
$\theta^2 (t) \simeq \frac{2(t-t_0)}{4 \Gamma^2 t_0}$
and 
$\bar{F}_{\rm GRB}^b |_{t_0} \simeq \frac{L_{\rm GRB}^b c 
\Gamma \Delta T}{4 \pi D_L^2 E^b \Delta^{\prime} (1+z)}$, 
where we have considered that the duration of prompt emission 
[$\sim \Gamma \Delta T/(1+z)$ in the comoving frame] is smeared as 
$\Delta^{\prime}/c \simeq \Gamma t/(1+z)$. 
The EIC suppression factor $f_{\rm EIC}$ is expressed as 
$f_{\rm EIC} = {\rm min}[1, \frac{(1+z)t_{\rm EIC}}{2 \Gamma \Delta T}]$, 
which becomes important if the EIC cooling is more relevant than the 
synchrotron and SSC cooling \citep[e.g.,][]{Wan+06}. 
The function $G(\varepsilon)$ represents the spectral shape of seed prompt 
photons with energy $\varepsilon$ in the comoving frame 
(e.g., $\varepsilon^b = (1+z) E^b/2 \Gamma$), where 
$G(\varepsilon)={(\varepsilon/\varepsilon^b)}^{-\beta_l+1}$ for 
$\varepsilon < \varepsilon^b$ and 
$G(\varepsilon)={(\varepsilon/\varepsilon^b)}^{-\beta_h+1}$ for 
$\varepsilon^b \leq \varepsilon$, respectively. 
The jet opening angle $\theta_j$ is also set to $\theta_j=0.1$ for both the prior 
and the main flows. 
As for the electron distribution of the prior flow, we exploit the standard external forward shock model \citep[e.g.,][]{SPN98} and adopt the following parameter set: 
$\mathcal{E}_k={10}^{52.5}$~erg, $n=1~{\rm cm}^{-3}$, $\epsilon_e=0.1$, $\epsilon_B=0.01$ and $p=2.4$. As for the prompt emission, the seed photon spectrum is assumed to be a broken power-law spectrum with $\beta_l = 1$ for $E<E^b$ and $\beta_h = 2.2$ for $E^b \leq E$. 
One might think that our approximation using a broken power law is not valid, since the prompt emission from some bursts such as GRB 090510 may have an extra component above $\sim 100$~MeV. However, even if we take into account this, our results are not changed because of the KN suppression at high energies. 
The high-energy cutoff due to the pair-creation is given by $E_{\rm}^{\rm cut} \sim 4~{\rm GeV}~{(L_{\rm GRB,51}^b)}^{-1} r_{i,14} \Gamma_{0,2.5}^4 {(1+z)}^{-1}$ \citep[e.g.,][]{LS01,GZ08,MI08} and we assume that the cutoff behavior as $1/[1+\tau_{\gamma \gamma} (E)]$ for simplicity \cite{Bar06}. The low-energy cutoff due to the synchrotron self-absorption is given by $E_{\rm}^{\rm sa} \sim 60~{\rm eV}~{(L_{\rm GRB,51}^b)}^{2/3} r_{i,14}^{-1} \Gamma_{0,2.5}^{-2/3} {(1+z)}^{-1}$ assuming that the radiative efficiency is order of unity \citep[e.g.,][]{RMZ04}. 
We consider the following two parameter sets for the other quantities. One expresses typical GRBs: 
$L_{\rm GRB}^b = {10}^{51.5}~{\rm erg} {\rm s}^{-1}$, 
$\Delta T/(1+z) = {10}^{1.5}$~s, $E^b=100$~keV, 
$\Gamma_0={10}^{2.5}$ and $r_i={10}^{14}$~cm. 
The other set expresses optically bright GRBs: 
$L_{\rm GRB}^b = {10}^{49.5}~{\rm erg} {\rm s}^{-1}$, 
$\Delta T/(1+z) = {10}^{1.5}$~s, $E^b=10$~eV, 
$\Gamma_0={10}^{2.5}$ and $r_i={10}^{15.5}$~cm. 
Such a case is also considered here, because bright optical emission was observed for some GRBs and the ratio of the optical flux to the gamma-ray flux $\mathcal{R} \equiv F_{\rm GRB, opt}/F_{\rm GRB, \gamma}$ is $\sim 10$ for GRB 041219A \cite{Ves+05} and $\sim 100$ for GRB 990123 \cite{Bri+99}.  
In order to demonstrate detectable cases, we hereafter show spectra and 
light curves only for $z=0.3$ with cosmological parameters 
$H_0=71~{\rm km}~{\rm s}^{-1}~{\rm Mpc}^{-1}$, $\Omega_M=0.3$, and 
$\Omega_\Lambda=0.7$. 

\begin{figure}
\includegraphics[width=\linewidth]{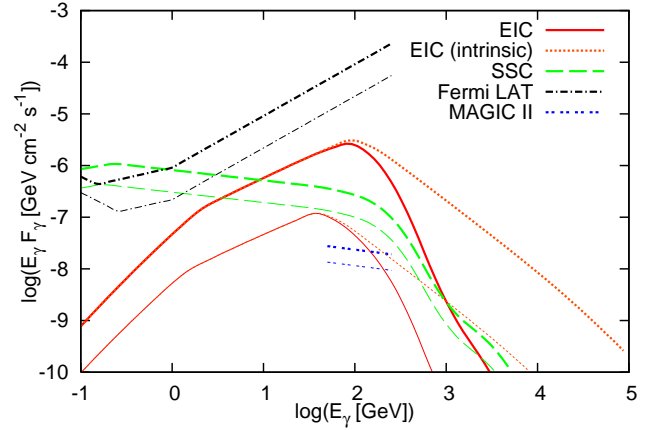}
\caption{Gamma-ray spectra of the EIC and SSC emission from GRB afterglows in the prior emission model for $t_0={10}^{2.5}$~s, plotted at $t={10}^{3}$ s (thick),  $t={10}^{3.5}$ s (thin), for the typical GRB with $L_{\rm GRB}^b = {10}^{51.5}~{\rm erg} {\rm s}^{-1}$, $\Delta T/(1+z) = {10}^{1.5}$~s, $E^b=100$~keV, $\Gamma_0={10}^{2.5}$ and $r_i={10}^{14}$~cm.  
The {\em Fermi}/LAT and MAGIC II sensitivities (with the duty factor of 20 \%) 
are also overlayed (Carmona et al. 2007). The sensitivity curves in the sky survey mode are used for the long time observations, although the possible continuous observations by LAT may 
improve the detectability by a factor of 3-5 (e.g., Gou \& M\'esz\'aros 2007).
}
\end{figure}
\begin{figure}
\includegraphics[width=\linewidth]{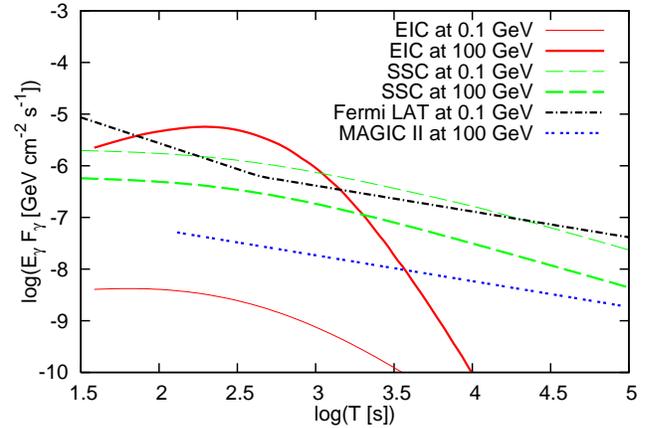}
\caption{
Gamma-ray light curves of the EIC and SSC emission from GRB afterglows in the prior emission model for $t_0={10}^{2.5}$~s, plotted at 0.1 GeV (thin) and 100 GeV (thick). 
The parameter set is the same as that used in Fig. 1.}
\end{figure}

Our results for the case of typical GRBs are shown in Figs.~1 and 2. As is expected, the EIC peak is located at $E_{\rm KN,1} \sim 10$--100~GeV. For this parameter set of $\epsilon_e/\epsilon_B = 10$ and $p=2.4$, the EIC flux is larger than the SSC flux above $\sim 10$~GeV
when $10^{2}~{\rm s}<T<10^{3.5}~{\rm s}$. The EIC flux is suppressed for a while just after the trigger time, because prompt photons interacting with relativistic electrons come from backwards ($\theta \sim 0$). Later, seed photons enter the prior flow with the angle $\theta \neq 0$, which contribute to the EIC flux. In Fig.~1, one can see that the KN suppression is prominent above 
$\sim 10$--100~GeV and the EIC peak is comparable to the SSC peak. The resulting spectra are basically explained by Eq.~(1). We can also see that the duration of the EIC emission is $\Delta t_{\rm EIC} \sim t_0={10}^{2.5}$~s, since the EIC emission lasts until we observe prompt photons entering the prior flow with $\theta \sim 1/\Gamma$. 
It is expected that \textit{Fermi} can detect the SSC afterglow emission at $\sim 0.1$--1~GeV. 
For closer and/or energetic bursts, we could even see the EIC bump in gamma-ray light curves at $\sim 10 $~GeV, 
In order to detect high-energy gamma rays with energy larger than 10~GeV, more suitable detectors are Cherenkov telescopes such as MAGIC, VERITAS and future Cherenkov Telescope Array (CTA). Although no firm signals have been hunted so far \cite{Abd+07,Alb+07,Aha+09}, fast follow-up observations are important to test the model. 
Note that the attenuation for the pair-creation process by the cosmic infrared background becomes important at such high energies. In all the figures we show, this attenuation is taken into account, 
using the low-IR model developed by Kneiske et al. 2004. Here, the pair echo emission is not included since it is beyond the scope of this work, although it could affect the observed afterglow emission if the intergalactic magnetic field in voids is weak enough \cite[e.g.,][]{RMZ04,Mur+09}. 

\begin{figure}
\includegraphics[width=\linewidth]{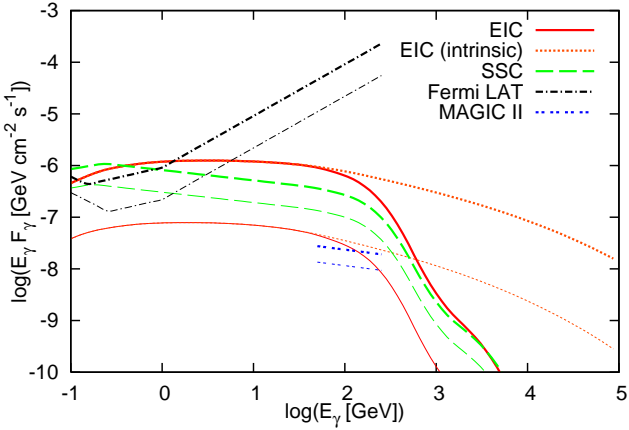}
\caption{The same as Fig. 1, but for the optically bright GRB with $L_{\rm GRB}^b = {10}^{49.5}~{\rm erg} {\rm s}^{-1}$, $\Delta T/(1+z) = {10}^{1.5}$~s, $E^b=10$~eV, $\Gamma_0={10}^{2.5}$ and $r_i={10}^{15.5}$~cm. 
}
\end{figure}
\begin{figure}
\includegraphics[width=\linewidth]{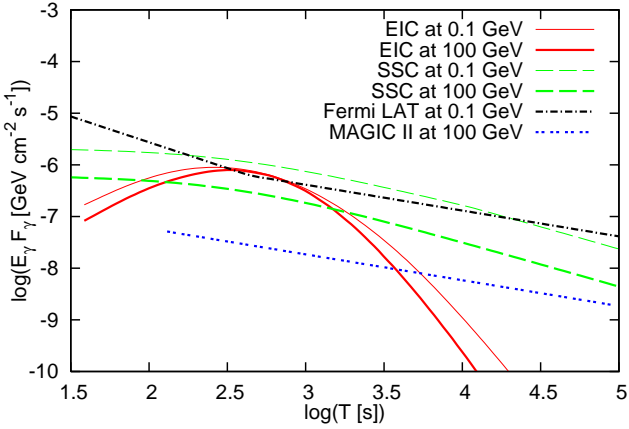}
\caption{
The same as Fig.1, but for the parameter set used in Fig. 3.}
\end{figure}

In Figs. 3 and 4, our results for a optically bright burst are shown. 
When copious optical/ultraviolet prompt photons are emitted from the main flow, we may observe the EIC emission dominant over the SSC emission above $\sim \rm GeV$.
The behavior of light curves are similar to those shown in Fig.~2, and we can see again that the duration of the EIC emission is $\Delta t_{\rm EIC} \sim t_0= {10}^{2.5}$~s. 
However, the resulting spectra are different from those shown in Fig.~1 because smaller values of $E^b \sim 10$~eV lead to the lower EIC peak of $\sim 0.1-1$~GeV. As is easily expected, the KN suppression is prominent above $\sim 10-100$~GeV, above which the EIC spectrum becomes steeper. In addition, in this case, the EIC cooling of electrons accelerated at the external shock is important, i.e., $\frac{2 \Gamma \Delta T}{(1+z) t_{\rm EIC}} > 1$, so that the resulting spectra can be modified from the simpler estimate neglecting the EIC cooling \citep{SE01,NAS09}. Furthermore, the afterglow emission itself would be affected by the EIC cooling and the SSC flux would also be suppressed. 
The EIC flux can dominate over the SSC flux above $\sim \rm GeV$. 
Hence, if the prior emission model is real, we should detect $\sim \rm GeV$ photons by \textit{Fermi} and $>$~a few $\times 10$~GeV photons by Cherenkov telescopes such as MAGIC, for such optically bright bursts. It also allows us to test the model.     

\section[]{Summary and Discussion}
In this work, we have studied the EIC emission predicted in the prior emission model and demonstrated that the EIC flux can be larger than the SSC flux at around the end time of the plateau phase above $\sim 10$~GeV. Although the situation depends on parameters, 
our results have suggested that observations of high-energy gamma rays provide important clues to testing the model in which the EIC emission is naturally expected, as well as observations at X-ray and optical bands \cite{LLZZ09}. 
The strategy for testing the model is as follows. 
First, we need to determine relevant afterglow parameters for the prior flow. When afterglows are well observed at X-ray and/or optical bands, the parameters such as $p$, $\epsilon_B$ and $\epsilon_e$ are determined in the context of the standard external forward shock theory\footnote{As long as the model is applicable. We may need to select bursts without chromatic breaks, since the model cannot explain them.}.
%
Furthermore, parameters on the prompt emission, such as $L_{\rm GRB}^b$ and $E^b$, can also be determined from observations, so that the EIC emission can be predicted.
Another uncertain parameter $r_i$ does not change our results for typical GRBs as long as $r_i < r$.
In particular, when $\epsilon_B/\epsilon_e$ and/or $p$ are large enough, the EIC flux becomes relatively larger than the SSC flux. Detections of characteristic features such as the EIC bump may support this model. 
The EIC peak is typically located at $\sim10$--100~GeV due to the KN effect. Therefore, ground-based gamma-ray observatories, such as MAGIC, VERITAS, High Energy Stereoscopic System Observatory (HESS), and CTA should be important, although very-high-energy gamma rays have not yet been observed so far for any GRBs \cite{Abd+07,Alb+07,Aha+09}. 
However, the detection rate of both the EIC and SSC emissions may not be large, because the rate of bursts occurring at within $z \sim 0.3$ is estimated as $\sim$~a few events per year \citep[e.g.,][]{LZVD07} and the detection rate also depends on several factors such as the detector sensitivity and field of view. 
%

We have calculated EIC spectra, assuming that both the jet opening angles and axes of the prior and main outflows are the same. As is discussed in Yamazaki (2009), it might not be the case to 
avoid the bright precursor emission. However, we can still expect the EIC emission. Even though the afterglow is dim in the early time, the prior outflow jet expands sideways, and the jet edge
crosses the line of sight, so that the prompt photons from the prior outflow come to the observer through the prior outflow (independently of the prompt emission mechanism).
In this case, the EIC emission occurs although the resulting flux is reduced by a factor of 2 
at most. 
However, if the jet axes of the two flows are different and the prompt emission occurs at $r_i \gg r$, then the EIC emission may not occur.   
 
The EIC emission studied in this work might be useful as a test for other two-component models. For example, the EIC emission can also be expected in the late prompt emission model \cite{Ghi+07}, which may be potentially important. 
In addition, high-energy gamma-ray afterglows might also be produced by hadronic mechanisms
\citep[for reviews, see][]{Mes06,Zha07,FP08}. This possibility is not in our focus, but it can be relevant if the baryon loading is large enough \citep[see e.g.,][and references there in]{PW05,Der07,Mur07}. Further complications may arise by flares, where high-energy emission can be expected via both the leptonic \cite{Wan+06,Wei+06,YD09} and hadronic mechanisms \cite{MN06}. 

Finally, let us briefly discuss the potential importance of the EIC emission on 
gamma-ray observations by \textit{Fermi}, although the detailed study is left as a future work.  
One of the observed characteristic properties is that high-energy ($\sim 1$--10~GeV) photons are delayed compared to $\sim \rm keV$--MeV photons \citep[see, e.g.,][]{TWM09}. 
In fact, we here point it out that EIC photons can be delayed by $\sim 10$--100~s when we consider the episode of ${\Delta t_{\rm EIC}} \sim t_0 \sim 10$--100~s. 
The strong EIC emission with the hard spectrum $E F_{\rm EIC} (E) \propto E^{1}$ below the EIC peak of $\sim 10$--100~GeV is also expected and detectable by \textit{Fermi} for extremely bright bursts (e.g., $L_{\rm GRB}^b \sim {10}^{54}~{\rm erg}~{\rm s}^{-1}$, $\mathcal{E}_k \sim {10}^{55}~{\rm erg}$ and $z \sim 1$). For sufficiently small $t_0$, the EIC process can be very efficient since small $r$ leads to large $Y_{\rm EIC}$, so that the EIC luminosity can roughly become the maximum, i.e., the afterglow electron luminosity, as $L_{\rm EIC} \sim L_e \sim \epsilon_e \mathcal{E}_k/t_0 \sim \epsilon_{e,-1} {10}^{52-53}~{\rm erg}~{\rm s}^{-1}$ which is one or two orders of magnitude smaller than $L_{\rm GRB}^b$. However, in this case, 
$t_0 \sim 10$--100~s is required, so that the prior emission model fails to explain the X-ray plateau phase lasting for $\sim {10}^{2-3}$~s.
Also, note that EIC light curves are essentially similar to those of high-latitude emission, so that they seem different from observed power-law behavior \cite{Ghi+09}.

\section*{Acknowledgments}
We thank B. Zhang and T. Nakamura for helpful comments. 
K.~M. and K.~T. acknowledge Grant-in-Aid from JSPS and NASA NNX08AL40G, 
respectively. This research was also supported by Grant-in-Aid from the
MEXT of Japan, nos.~21740184 (R.~Y.), ~19104006,~19740139,~21105509 (S.~N.),~21684014,~18740147, (K.~I.), and~19047004.


\bsp

\label{lastpage}

\end{document}